\documentclass{ws-procs11x85}
\usepackage{balance}
\usepackage{epsfig}
\def\Journal#1#2#3#4{{#1} {\bf #2}, #3 (#4)}

\makeindex

\begin{document}

\title{Ordering Algorithms and Confidence Intervals in
the Presence of Nuisance Parameters}

\author{Giovanni Punzi}

\address{I.N.F.N.-Sezione di Pisa, Largo B. Pontecorvo 3,
56100 Pisa, Italy\\ E-mail: giovanni.punzi@pi.infn.it}

%%%%%%%%%%%%%%%%%%%%%%%%%%%%%%%%%%%%%%%%%%%%%%%%%%%%%%%%%%%%%%%%%%%%%%%%%
% You may repeat \author \address as often as necessary                 %
%%%%%%%%%%%%%%%%%%%%%%%%%%%%%%%%%%%%%%%%%%%%%%%%%%%%%%%%%%%%%%%%%%%%%%%%%

\twocolumn[\maketitle\begin{abstract} We discuss some issues arising
in the evaluation of confidence intervals in the presence of nuisance
parameters (systematic uncertainties) by means of direct Neyman
construction in multi--dimensional space.  While this kind of procedure
provides rigorous coverage, it may be affected by large overcoverage,
and/or produce results with counterintuitive behavior with respect to
the uncertainty on the nuisance parameters, or other undesirable
properties.  We describe a choice of ordering algorithm
that provides results with good general properties, the correct
behavior for small uncertainties, and limited overcoverage.
\end{abstract}]

\bodymatter

\baselineskip=13.07pt
\section{Introduction}

%%%%%%%%%%%%%%%%%%%%%%%%%%%%%%%%%%%%%%%%%%%%%%%%
\sloppy

A conceptually straightforward method to incorporate
systematics into Confidence Limits is to apply
the usual Neyman construction directly on the complete {\em pdf} of the 
problem, including the set of additional 
parameters $\nu$ describing the systematic effects,
and then project the solution on the space of parameters of interest $\mu$.
Systematic uncertainties may take the form of an allowed range for 
the $\nu$'s, or may be defined by the observables of the problem. 
Although the method can be applied to a more general situation, we will assume 
in the following discussion that measurements are available of some
(``subsidiary'') observable(s) $y$, whose only purpose is
to provide information on the systematic parameters, through the 
dependence of their {\em pdf} on $\nu$.
In this case, one will consider the overall {\em pdf}:
\begin{equation}
    p((x,y)|(\mu,\nu))
\end{equation}
that gives the joint probability of observing the value of the ``physics
observables" $x$ plus all ``systematic measurements" $y$, given all
unknown parameters, physics and systematics.
One starts by deriving Confidence Limits in the larger $(\mu,\nu)$ space
from the observed values of $(x,y)$ with the same procedure that could have been used in 
absence of systematics to derive limits on $\mu$:
one simply needs to sample a number of points inside the parameter space and require 
coverage for each of them.
Then, in order to get results containing
only the physical parameters, one needs to project the confidence
region in $(\mu,\nu)$ onto the $\mu$ space, so as to get rid of 
unwanted information on the nuisance parameters.

Although the above procedure is general, conceptually simple, and rigorous, 
other methods have been preferred in the vast majority of 
problems in physics. This can be ascribed to a few important 
difficulties with this method. 
To begin with, the problem of numerical calculation of Confidence Regions (CRs) in multi--dimensional spaces 
is often quite complex and CPU--consuming.
Then there is a non--trivial question of what ordering algorithm to 
use in the Neyman construction. There is an issue of ``efficiency'', 
or power, of the solution, because projecting the 
band on the $\mu$ space effectively means to inflate a
limited region in $(\mu,\nu)$ to an unlimited band in the $\nu$ 
direction, 
thereby increasing the coverage for all additional points $(\mu,\nu)$ 
included. This means that the final limits quoted on $\mu$ will almost 
always {\em overcover}, and sometimes badly, especially when $\nu$ 
has many dimensions; this is indeed the case with 
standard choices of ordering\cite{Durham}.
A related additional problem is that the behavior of the limits when 
the systematic uncertainty approaches zero is in many cases unsatisfying. It often happens 
that the limit for small systematics is quite different from the result one 
would quote in absence of that systematic; this problem, however, 
is not unique to the projection method.

If the above problems could be alleviated, this methodology could find
greater use in HEP. 

\section{A Benchmark problem}\label{sec:problem}

Our discussion, although general, will be centered on a specific 
problem that has been the initial motivation for this work: a Poisson 
distributed signal in presence of a known background, with a systematic 
uncertainty on the signal normalization (efficiency).
We have:
\begin{eqnarray}
    x\sim Pois(\epsilon \mu + b) \mbox{\hspace{0.5cm}} & , \mbox{\hspace{0.5cm}}
    e\sim G(\epsilon,\sigma)
    \end{eqnarray}
where $e$ is the result of a subsidiary measurement
with resolution $\sigma$ of the unknown efficiency $\epsilon$, 
which is intended to be a generic 
``normalization factor'', not necessarily smaller than one.
In the following we will mostly assume a normal distribution for $G$ 
for simplicity; the possibility of
negative values of the efficiency estimate does not pose any problems
to the algorithms 
discussed in this document. This can actually occur, for 
instance, when the efficiency measurement implies some sort of
background--subtraction procedure\footnote{This simple and common example has been selected by the CDF 
statistics committee as a benchmark in performing comparisons between a number of 
different methods. A minor difference from the current example is that a positive, Poisson--like distribution is 
assumed for the subsidiary measurement instead of a Gaussian, in order to avoid problems 
with Bayesian treatment\cite{CDF-Bayesian}.}.

%%%%%%%%%%%%%%%%%%%%%%%%%%%%%%%%%%%%%%%%%%%%%%%%%
\section{Looking for an optimal band}

What one would like to accomplish is to find a clever 
enough rule for constructing the initial Confidence Band, to minimize the amount 
of unnecessary coverage added when the band 
is projected onto the ``interesting parameters'' space. 
It is not obvious what the minimum is for a particular problem, because the frequentist requirement of 
minimum coverage for every possible true value of the 
parameters may imply some minimum amount of overcoverage, which is 
unavoidable regardless 
of the algorithm used in the construction, much in the way overcoverage occurs in discrete 
problems. Therefore, there is no reason for being a--priori discouraged about 
the capability of the projection method to provide powerful solutions 
(that is, narrow intervals). A striking demonstration of this is 
provided by the use of the projection method, with an appropriately designed algorithm for band 
construction, in producing a more efficient solution to a classical, 
well--explored problem like the ratio of Poisson means\cite{Cousins-projection}.

It is intuitively obvious that in order to obtain an efficient 
solution, the initial confidence band must extend as far as 
possible along the direction of the nuisance parameter. This is 
not trivial to achieve, since the band needs to be built in the $(x,e)$ 
space, while the objective is to produce a desired shape in the 
$(\mu,\epsilon)$ space. 
A good general requirement to impose is that, given any two sections 
of the band at two fixed 
values of the nuisance parameter $\epsilon$, one must be completely 
included in the other. It is intuitive that a band cannot be optimal 
if it does not satisfy this requirement, because if one had to take one 
of the two sections and expand it to completely include the other, 
the projected confidence region in $\mu$ would be unaffected, and 
conversely one could exploit the coverage gained in this way to trim 
a part of the exceeding part of the chosen section, 
thus creating the conditions for a tightening of 
the projected confidence region.

\section{Ordering Algorithm}

One way to define how to construct the confidence band in the complete 
space is to derive it from an ordering function $f(x,e;\mu,\epsilon)$, 
so that the confidence band is defined by the 
inequality $f(x,e;\mu,\epsilon) > c(\mu,\epsilon)$, where the 
threshold $c$ is determined for each value of the parameters from the 
usual Neyman's requirement of coverage: 
\begin{equation}
    \int_{f(x,e;\mu,\epsilon) > c(\mu,\epsilon)}{p(x,e|\mu,\epsilon) dx 
 de} \geq CL
 \end{equation}
 
where $CL$ is the desired Confidence Level.
It is worth noting that this is not the only conceivable way to 
define a band satisfying the coverage condition\cite{Cousins-projection},
but it is attractive for reasons of simplicity.
A simple way to implement in an ordering 
algorithm the requirement of 
inclusion formulated in the previous section is to impose that 
$f(x,e;\mu,\epsilon)$ is 
independent of $\epsilon$: 
$f(x,e;\mu,\epsilon_{1})=f(x,e;\mu,\epsilon_{2})$. In this way, sections 
taken at different $\epsilon$ for the 
same value of $\mu$ will only differ in the value of 
$c(\mu,\epsilon)$, and will therefore be included in one another.
This requirement is also very convenient from the point of view of 
computing, as it implies that the ordering function $f$ need only be 
calculated once for every $\mu$.

\begin{figure}[htb]
\flushleft
{\bf a)}
\vspace{-0.5cm}
\center
\includegraphics[width=\columnwidth]{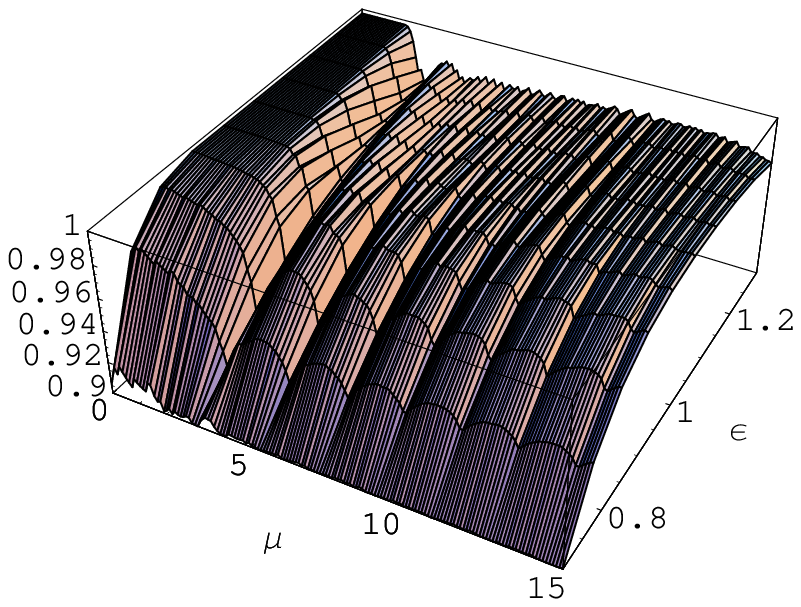}
\vspace{-1cm}
\flushleft
{\bf b)}
\vspace{-0.5cm}
\center
\includegraphics[width=\columnwidth]{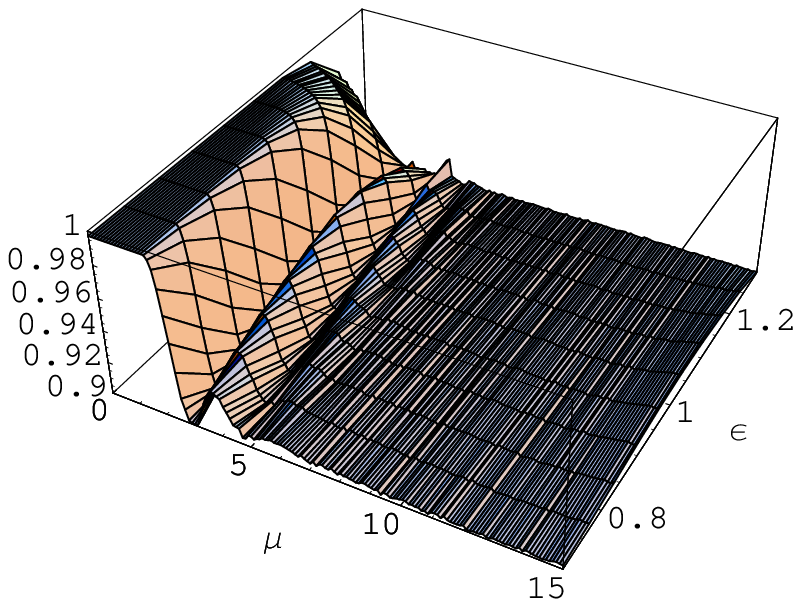}
\caption{\label{fig:upper}\label{fig:naiveupper} Coverage plots for 
upper limits with a naive ordering (a), and with the ordering of 
eq.(\ref{eq:ordering}) (b). The efficiency is measured with a Gaussian 
uncertainty, $\sigma=0.1$.}
\end{figure}

As an additional requirement, we want the 
projected confidence regions to converge to the results in absence of 
systematic uncertainty when the size of the uncertainty goes to zero. 
We do not restrict to a specific ordering (one may want to be able to 
choose, for 
instance, between central and upper 
limits), so we start from a given generic ordering function 
$f_{0}(x;\mu)$ in the restricted space. 
This defines the behavior of the ordering function 
along the direction of observable $x$, but careless extension of any 
such rule to the whole $(x,e)$ space will not work. As an example, extending
the trivial ordering used to achieve upper limits 
($f_{0}(x;\mu)=x$) results in substantial overcoverage 
(see fig.\ref{fig:naiveupper}a).
We need additional criteria to ensure proper behavior in the 
subsidiary observable $e$. We don't want to give special 
preference to any values, because this will amount to attempting to 
extract information on the nuisance parameter, while we want to 
maximize information on the physical parameter $\mu$.
We do this by choosing the following ordering function:
\begin{equation}\label{eq:ordering}
    f(x,e;\mu) = 
    \int_{f_{0}(x')<f_{0}(x)}{p(x'|e;\mu,\hat{\epsilon}(e)) dx'}
\end{equation}

where $\hat{\epsilon}(e)$ is the maximum--Likelihood estimate of $\epsilon$ for the 
given $e$. 
That implies that the same integrated conditional probability will be 
contained in the band for each value of $e$. 

We make an exception to the rule of being indifferent to the value of 
$e,\epsilon$ for very unlikely values: we select an interval of values 
$[e_{min},e_{max}]$ such that the probability for a measurement to fall 
outside is $\ll 1-CL$, and assign lowest rank to all points lying 
outside this interval. From the above conditions, they will never be
reached by the ordering procedure, so they can simply be ignored, 
which saves computation. This clipping technique has already been advocated
as a help in keeping the projections small\cite{Kyle03}; in our 
context however it seemed to have no significant effects beyond saving 
computation.

\section{Results}
We have applied the ordering rule of equation (\ref{eq:ordering}) to 
our problem of choice (sec. \ref{sec:problem}), with an ordering 
$f_{0}$ 
corresponding to upper limits.
Fig.~\ref{fig:upper}b shows that this time very little
overcoverage is obtained, except from some discretization--related 
``ripples''. 
It is interesting to note that these limits are tighter than the limits obtained 
with other popular methods (compare, for instance, the coverage 
obtained for the same problem with Bayesian\cite{CDF-Bayesian} or
Cousins--Highland methods\cite{CDF-CH,T-Conrad}),
although guaranteed by construction to cover for every possible value
of both $\mu$ and $\epsilon$. 
This confirms the capability of the projection method to produce powerful results,
when used in conjunction with an appropriate ordering 
algorithm, as per Eq.~(\ref{eq:ordering}). 

\begin{figure}[htb]
\center
\includegraphics[width=\columnwidth]{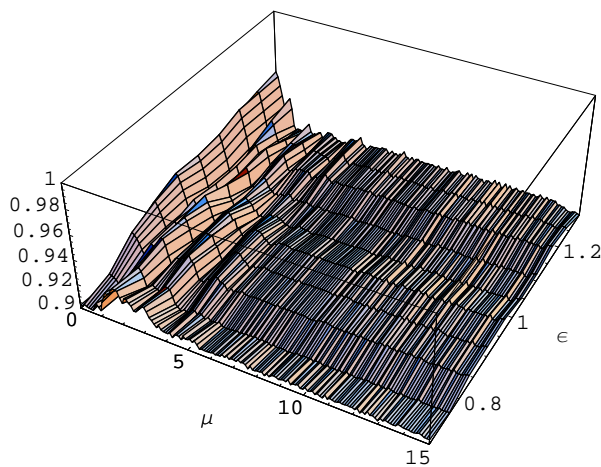}
\caption{\label{fig:FC} Coverage plot for Unified limits, 
Gaussian uncertainty, $b=3,\sigma=0.1$.}
\end{figure}

The procedure we have described can be used with any other desired
ordering.
If we apply it to Unified Intervals\cite{FC}, we find an interesting 
fact: because of the Likelihood Ratio theorem, ensuring the independence of the 
distribution from true parameter values, the ordering 
algorithm defined by Eq.~(\ref{eq:ordering}) is approximately 
equivalent to ordering based on the ratio of profile Likelihoods.
% \begin{equation}\label{eq:prof-LR
%   
%     \end{equation}
That quantity has been suggested as a good intuitive ordering to 
use in handling systematics since \cite{Kendall}, and has been used 
in neutrino experiments\cite{Durham,CHOOZ} (with a
conditional frequentist motivation),
and in a problem very similar to ours, the Poisson with uncertainty on 
background\cite{Kyle03}. 
It reappears here as an approximation of the more 
general rule defined by Eq.~(\ref{eq:ordering}). 
Fig~\ref{fig:FC} shows that coverage plot for our benchmark problem, 
which is close to the nominal constant $0.9$, 
indicating that there is very little to be further gained.
%Sample results are reported in Tab.~\ref{tab:limits}
\section{Continuity}

\begin{figure}[htb]
\center
\includegraphics[width=\columnwidth]{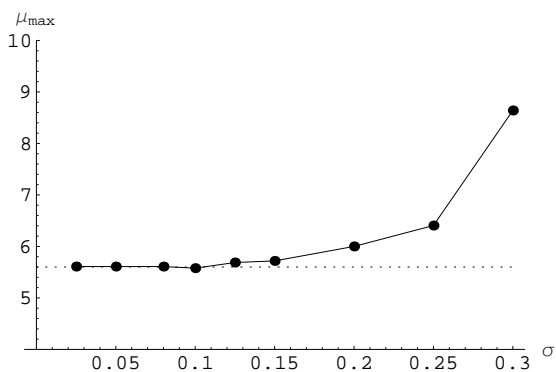}
\caption{\label{fig:0-syst-fixed} Behavior of upper limit when 
$\sigma \to 0$, same problem as in Fig.~\ref{fig:FC}.}
\end{figure}

One of our initial goals was to obtain a continuous 
behavior when $\sigma_{syst}\to 0$. In previous examples, 
although the limit is approximated 
much better than with other frequentist methods (see for 
instance\cite{Rolke}), there is still a slight difference. For 
instance, the upper limit with the Unified method at $90\%$ for $n=4, 
b=3$ is\cite{FC} $5.6$ , while our results approach $\simeq 5.47$ 
when $\sigma \to 0$. More annoyingly, the limit found with 
systematics is lower. This is a well 
known problem, tied to the transition between 
discrete and continuous regime\cite{Feldman-fnal}, and is pretty much 
independent of the specific algorithm. However, our method for 
evaluating limits allows a very simple fix, requiring 
no alterations to the ordering: all that is needed is to keep the size of 
the grid used in the numerical calculations from becoming too small
in the direction of the nuisance observable. 
This has a natural justification under 
the same principles that guided the general design of our 
algorithm: we are trying to disregard detailed information on the  
the subsidiary observable, in favor of information on the physics parameter $\mu$. 
In our problem, by choosing a minimum step $\Delta e = 0.1$ we 
obtain perfect continuity at zero (fig.~\ref{fig:0-syst-fixed}). A 
side effect of this limitation is to save some computing time.

\section{Systematic uncertainties given as ranges}

The approach we have described has wider application than the 
examples mentioned above. For instance, it can handle in a 
natural way the important situations in which no subsidiary measurement in available to 
provide information on the nuisance parameter. This often occurs in 
real life: the systematic uncertainty may be due to a physical 
constraint, or related to a choice 
within a range (discrete or continuous) of theoretical 
predictions or assumptions, or can 
otherwise be specified in a way that is not detailed enough to 
uniquely identify a probability distribution.
In these cases, usually the only available information on $\epsilon$ 
is represented by a range of admissible values. 

This situation is automatically handled by our approach: 
one simply has one less observable to worry about, but the rest of 
the construction works exactly in the same way.
In fact, calculations are much faster with the lack of a subsidiary 
measurement, so that when dealing with small systematics 
it is actually more convenient to transform any possible nuisance measurement into 
an appropriate range for the nuisance parameter, and simply use that 
information as input, in order to save computing time.
Again, our tests yielded very limited 
overcoverage, compatible with what was required simply by the 
discrete nature of the problem.

It is worth noting that a range of values is 
not at all equivalent to a uniform distribution, which 
implies more precise knowledge. For instance, by comparing the limits 
obtained in the two cases, it is seen that the limits for the range case
are looser then in the uniform distribution case, as intuitively expected
due to the smaller information content in a statement about a range 
(see Table~\ref{tab:limits}).
This is in contrast with what happens in a Bayesian approach, 
where a prior function is always required, and a uniform distribution 
is often chosen to represent lack of information.

In general, treating systematic uncertainties as ranges is a good 
candidate approach to problems with many 
nuisance parameters, as it allows big savings in CPU time, in 
addition to avoiding the 
trouble of having to worry about the accuracy of the distributions 
assumed to represent the systematic uncertainties.


\begin{table*}[htb]
 \tbl{\label{tab:limits} Confidence Limits for Poisson+background 
 with systematic uncertainty on the efficiency,
 obtained by extending Unified Limits through Eq.(\ref{eq:ordering}).
 Results are given for $b=3$, $e=1.0$, and various 
 models of uncertainty on $\epsilon$ (see text).}
{ \centering
% \tiny
 \begin{tabular}{|c||l|l|l|l|}
        \hline
          &    Without    & Gaussian     &  Uniform    & Range  \\
$n_{obs}$ & systematics & $\sigma=0.1$ & $\pm 0.15$ & $\pm 0.15$\\
        \hline\hline
         0 & 0.00 ,  1.08 & 0.0 ,  1.1 & 0.0 ,  0.9 & 0.0 ,  1.0\\
         1 & 0.00 ,  1.88 & 0,0 ,  1.9 & 0.0 ,  1.7 & 0.0 ,  1.9\\
         2 & 0.00 ,  3.04 & 0.0 ,  3.0 & 0.0 ,  2.7 & 0.0 ,  3.0\\
         3 & 0.00 ,  4.42 & 0.0 ,  4.4 & 0.0 ,  4.0 & 0.0 ,  4.5\\
         4 & 0.00 ,  5.60 & 0.0 ,  5.9 & 0.0 ,  5.4 & 0.0 ,  6.0\\
         5 & 0.00 ,  6.99 & 0.0 ,  7.4 & 0.0 ,  6.9 & 0.0 ,  7.4\\
         6 & 0.15 ,  8.47 & 0.0 ,  8.9 & 0.2 ,  8.2 & 0.1 ,  8.9\\
         7 & 0.89 ,  9.53 & 0.9 , 10.3 & 1.0 ,  9.6 & 0.8 , 10.4\\
         8 & 1.51 , 10.99 & 1.4 , 11.7 & 1.5 , 10.9 & 1.3 , 11.8\\
         9 & 1.88 , 12.30 & 2.0 , 13.1 & 2.1 , 12.3 & 1.9 , 13.1\\
         \hline
     \end{tabular}
     }
 \end{table*}

\section{Conclusions}

We have presented a general method to incorporate systematic uncertainties in a 
limit calculation in a rigorous frequentist way, which is powerful (does
not produce large overcoverage), has the right limit for small uncertainties, 
can be used even with uncertainties given as ranges, 
and can easily be calculated in practice. This is based on 
projection of a traditional Neyman construction with an ordering algorithm 
specified by Eq.~(\ref{eq:ordering}). We have applied it to the 
specific problem of Poisson measurement with an uncertainty on the efficiency.

\section*{Acknowledgments}

The Author is deeply indebted to Louis Lyons for continuing inspirational interaction 
and for organizing such pleasant and productive meetings; also wishes to 
thank Robert Cousins for helpful comments, and all members of the CDF statistics
committee for many stimulating discussions.

\end{document}